\documentclass[12pt, preprint]{aastex}

\shorttitle{VERITAS Observations of PG 1553+113}
\shortauthors{Orr et al.}

\begin{document}

\title{VERITAS Observations of the BL Lac Object PG 1553+113}

\author{
E.~Aliu\altaffilmark{1},
A.~Archer\altaffilmark{2},
T.~Aune\altaffilmark{3},
A.~Barnacka\altaffilmark{4},
B.~Behera\altaffilmark{5},
M.~Beilicke\altaffilmark{2},
W.~Benbow\altaffilmark{6},
K.~Berger\altaffilmark{7},
R.~Bird\altaffilmark{8},
J.~H.~Buckley\altaffilmark{2},
V.~Bugaev\altaffilmark{2},
K.~Byrum\altaffilmark{9},
J.~V~Cardenzana\altaffilmark{10},
M.~Cerruti\altaffilmark{6},
X.~Chen\altaffilmark{11,5},
L.~Ciupik\altaffilmark{12},
M.~P.~Connolly\altaffilmark{13},
W.~Cui\altaffilmark{14},
H.~J.~Dickinson\altaffilmark{10},
J.~Dumm\altaffilmark{15},
J.~D.~Eisch\altaffilmark{10},
M.~Errando\altaffilmark{1},
A.~Falcone\altaffilmark{16},
S.~Federici\altaffilmark{5,11},
Q.~Feng\altaffilmark{14},
J.~P.~Finley\altaffilmark{14},
P.~Fortin\altaffilmark{6},
L.~Fortson\altaffilmark{15},
A.~Furniss\altaffilmark{17},
N.~Galante\altaffilmark{6},
G.~H.~Gillanders\altaffilmark{13},
S.~Griffin\altaffilmark{18},
S.~T.~Griffiths\altaffilmark{19},
J.~Grube\altaffilmark{12},
G.~Gyuk\altaffilmark{12},
N.~H{\aa}kansson\altaffilmark{11},
D.~Hanna\altaffilmark{18},
J.~Holder\altaffilmark{7},
G.~Hughes\altaffilmark{5},
T.~B.~Humensky\altaffilmark{20},
C.~A.~Johnson\altaffilmark{17},
P.~Kaaret\altaffilmark{19},
P.~Kar\altaffilmark{21},
M.~Kertzman\altaffilmark{22},
Y.~Khassen\altaffilmark{8},
D.~Kieda\altaffilmark{21},
H.~Krawczynski\altaffilmark{2},
F.~Krennrich\altaffilmark{10},
S.~Kumar\altaffilmark{7},
M.~J.~Lang\altaffilmark{13},
A.~Madhavan\altaffilmark{10},
S.~McArthur\altaffilmark{23},
A.~McCann\altaffilmark{24},
K.~Meagher\altaffilmark{25},
J.~Millis\altaffilmark{26},
P.~Moriarty\altaffilmark{27,13},
D.~Nieto\altaffilmark{20},
A.~O'Faol\'{a}in de Bhr\'{o}ithe\altaffilmark{5},
R.~A.~Ong\altaffilmark{3},
M.~Orr\altaffilmark{10},
A.~N.~Otte\altaffilmark{25},
N.~Park\altaffilmark{23},
J.~S.~Perkins\altaffilmark{28},
M.~Pohl\altaffilmark{11,5},
A.~Popkow\altaffilmark{3},
H.~Prokoph\altaffilmark{5},
E.~Pueschel\altaffilmark{8},
J.~Quinn\altaffilmark{8},
K.~Ragan\altaffilmark{18},
J.~Rajotte\altaffilmark{18},
L.~C.~Reyes\altaffilmark{29},
P.~T.~Reynolds\altaffilmark{30},
G.~T.~Richards\altaffilmark{25},
E.~Roache\altaffilmark{6},
G.~H.~Sembroski\altaffilmark{14},
K.~Shahinyan\altaffilmark{15},
D.~Staszak\altaffilmark{18},
I.~Telezhinsky\altaffilmark{11,5},
J.~V.~Tucci\altaffilmark{14},
J.~Tyler\altaffilmark{18},
A.~Varlotta\altaffilmark{14},
V.~V.~Vassiliev\altaffilmark{3},
S.~P.~Wakely\altaffilmark{23},
A.~Weinstein\altaffilmark{10},
R.~Welsing\altaffilmark{5},
A.~Wilhelm\altaffilmark{11,5},
D.~A.~Williams\altaffilmark{17},
B.~Zitzer\altaffilmark{9}
}

\altaffiltext{1}{Department of Physics and Astronomy, Barnard College, Columbia University, NY 10027, USA}
\altaffiltext{2}{Department of Physics, Washington University, St. Louis, MO 63130, USA}
\altaffiltext{3}{Department of Physics and Astronomy, University of California, Los Angeles, CA 90095, USA}
\altaffiltext{4}{Harvard-Smithsonian Center for Astrophysics, 60 Garden Street, Cambridge, MA 02138, USA}
\altaffiltext{5}{DESY, Platanenallee 6, 15738 Zeuthen, Germany}
\altaffiltext{6}{Fred Lawrence Whipple Observatory, Harvard-Smithsonian Center for Astrophysics, Amado, AZ 85645, USA}
\altaffiltext{7}{Department of Physics and Astronomy and the Bartol Research Institute, University of Delaware, Newark, DE 19716, USA}
\altaffiltext{8}{School of Physics, University College Dublin, Belfield, Dublin 4, Ireland}
\altaffiltext{9}{Argonne National Laboratory, 9700 S. Cass Avenue, Argonne, IL 60439, USA}
\altaffiltext{10}{Department of Physics and Astronomy, Iowa State University, Ames, IA 50011, USA}
\altaffiltext{11}{Institute of Physics and Astronomy, University of Potsdam, 14476 Potsdam-Golm, Germany}
\altaffiltext{12}{Astronomy Department, Adler Planetarium and Astronomy Museum, Chicago, IL 60605, USA}
\altaffiltext{13}{School of Physics, National University of Ireland Galway, University Road, Galway, Ireland}
\altaffiltext{14}{Department of Physics and Astronomy, Purdue University, West Lafayette, IN 47907, USA}
\altaffiltext{15}{School of Physics and Astronomy, University of Minnesota, Minneapolis, MN 55455, USA}
\altaffiltext{16}{Department of Astronomy and Astrophysics, 525 Davey Lab, Pennsylvania State University, University Park, PA 16802, USA}
\altaffiltext{17}{Santa Cruz Institute for Particle Physics and Department of Physics, University of California, Santa Cruz, CA 95064, USA}
\altaffiltext{18}{Physics Department, McGill University, Montreal, QC H3A 2T8, Canada}
\altaffiltext{19}{Department of Physics and Astronomy, University of Iowa, Van Allen Hall, Iowa City, IA 52242, USA}
\altaffiltext{20}{Physics Department, Columbia University, New York, NY 10027, USA}
\altaffiltext{21}{Department of Physics and Astronomy, University of Utah, Salt Lake City, UT 84112, USA}
\altaffiltext{22}{Department of Physics and Astronomy, DePauw University, Greencastle, IN 46135-0037, USA}
\altaffiltext{23}{Enrico Fermi Institute, University of Chicago, Chicago, IL 60637, USA}
\altaffiltext{24}{Kavli Institute for Cosmological Physics, University of Chicago, Chicago, IL 60637, USA}
\altaffiltext{25}{School of Physics and Center for Relativistic Astrophysics, Georgia Institute of Technology, 837 State Street NW, Atlanta, GA 30332-0430}
\altaffiltext{26}{Department of Physics, Anderson University, 1100 East 5th Street, Anderson, IN 46012}
\altaffiltext{27}{Department of Life and Physical Sciences, Galway-Mayo Institute of Technology, Dublin Road, Galway, Ireland}
\altaffiltext{28}{N.A.S.A./Goddard Space-Flight Center, Code 661, Greenbelt, MD 20771, USA}
\altaffiltext{29}{Physics Department, California Polytechnic State University, San Luis Obispo, CA 94307, USA}
\altaffiltext{30}{Department of Applied Physics and Instrumentation, Cork Institute of Technology, Bishopstown, Cork, Ireland}

\begin{abstract}
We present results from VERITAS observations of the BL Lac object PG 1553+113 spanning the years 2010, 2011, and 2012.  
The time-averaged spectrum, measured between 160 and 560\,GeV, is well described by a power law with a spectral index of $4.33 \pm 0.09$.  The time-averaged integral flux above $200\,$GeV measured for this period was $(1.69 \pm 0.06) \times 10^{-11} \, \mathrm{ph} \, \mathrm{cm}^{-2} \, \mathrm{s}^{-1}$, corresponding to 6.9\% of the Crab Nebula flux.  We also present the combined $\gamma$-ray spectrum from the Fermi Large Area Telescope and VERITAS covering an energy range from 100~MeV to 560~GeV.  The data are well fit by a power law with an exponential cutoff at  $\rm  {101.9 \pm 3.2  \, \mathrm{GeV}} $.   The origin of the cutoff could be intrinsic to PG~1553+113 or be due to the  $\gamma$-ray opacity of our universe through pair production off  the extragalactic background light (EBL).   Given lower limits to the redshift of  $\rm z \negthinspace >  \negthinspace  0.395$ based on optical/UV observations of PG~1553+113, the cutoff would be dominated by EBL absorption.    Conversely, the small statistical uncertainties of the VERITAS energy spectrum have allowed us to provide a robust upper limit on the redshift of PG 1553+113 of $z \negthinspace \leq \negthinspace 0.62$. 
A strongly-elevated mean flux of $(2.50 \pm 0.14) \times 10^{-11} \, \mathrm{ph} \, \mathrm{cm}^{-2} \, \mathrm{s}^{-1}$ (10.3\% of the Crab Nebula flux) was observed during 2012, with the daily flux reaching as high as $(4.44 \pm 0.71) \times 10^{-11} \, \mathrm{ph} \, \mathrm{cm}^{-2} \, \mathrm{s}^{-1}$ (18.3\% of the Crab Nebula flux) on MJD 56048.  The light curve measured during the 2012 observing season is marginally inconsistent with a steady flux,  giving a $\chi^2$ probability for a steady flux of 0.03\%.  
\end{abstract}

\keywords{BL Lac objects: general - BL Lacertae objects: individual  (PG 1553+113  = VER J1555+111)}

\section{Introduction}
\label{sec:Intro}
PG 1553+113 was first discovered by \citet{Green:1986} and is classified as a high-frequency-peaked BL Lac  object (HBL) \citep{Falomo:1994,Giommi:1995,Beckmann:2002}.  Evidence of very-high-energy (VHE; $\rm E \ge 100$~GeV) gamma-ray emission from this source was first reported by H.E.S.S. in 2005 \citep{Aharonian:2006} and was later confirmed by observations made with the MAGIC telescope in 2005 and 2006 \citep{Albert:2007}.  Due to its featureless optical spectrum, the redshift of PG 1553+113 remains highly uncertain.  Constraints on the redshift, however, have been continually narrowing with improved optical measurements and limits from VHE observations  (e.g., \citet{Sbarufatti:2006,Treves:2007,Aharonian:2006,Mazin:2007}).  Recent measurements using the \textit{Hubble Space Telescope} Cosmic Origins Spectrograph (COS) have yielded the strictest redshift constraints to date -- setting a firm lower limit of $z \negthinspace > \negthinspace 0.395$ \citep{Danforth:2010}.

PG 1553+113 is readily detected in the high-energy (HE; $\sim \negthinspace 100\,$MeV to $100\,$GeV) and VHE gamma-ray regimes.  The Large Area Telescope (LAT), on board the \textit{Fermi} satellite, obtained a detection for this source after the first three months of observations, with a significance greater than 30 standard deviations ($\sigma$) above the background \citep{Abdo:2009}. The imaging atmospheric Cherenkov telescope array, VERITAS, is capable of detecting PG 1553+113 above 100\,GeV with a significance of $5\sigma$ after $\sim \negthinspace 43$ minutes of observations, given its average flux.  Previous measurements of PG 1553+113, made by H.E.S.S., yielded a time-averaged VHE spectral index $\rm \Gamma$ (refers to photon spectrum $\rm dN/dE \propto E^{-\Gamma}$)~of~$4.46 \pm 0.34$ between $225\,$GeV and $1.3\,$TeV \citep{Aharonian:2008}, consistent with measurements by MAGIC \citep{Albert:2007}.  The \textit{Fermi}-LAT 2 Year Catalog reports the spectral index between $100\,$ MeV and $100\,$GeV to be $1.67 \pm 0.02$ \citep{Nolan:2012}.\footnote{The full two year \textit{Fermi}-LAT catalog can be found online at: \\ \texttt{http://fermi.gsfc.nasa.gov/ssc/data/access/lat/2yr\_catalog/}}

The results from VERITAS observations of PG 1553+113 presented in this paper are organized as follows:  Sections \ref{sec:VERITASObs} and \ref{sec:FermiObs} summarize the VERITAS and \textit{Fermi}-LAT observations of PG 1553+113, respectively.  Source variability at high- and very-high energies is discussed in Section \ref{sec:Variability}.  Constraints on the source redshift obtained using VERITAS observations are presented alongside previous constraints in Section \ref{sec:Redshift}.  Finally, a discussion and conclusions are given in Section \ref{sec:Discussion}.

\section{VERITAS Observations}
\label{sec:VERITASObs}
The VERITAS array of imaging atmospheric Cherenkov telescopes (IACTs) is located in southern Arizona (31$^{\circ}$40'30"N, 110$^\circ$57'07"W) at an elevation of $1.3\,$km above sea level and is described in \cite{Kieda:2013}.  The array is comprised of four Davies-Cotton telescopes, each 12$\,$m in diameter, arranged in an approximate diamond configuration with telescope separations of $\sim \negthinspace 100\,$m.  The optical system of each telescope has a focal length of 12$\,$m, and consists of  345 individual hexagonal mirror facets with a total mirror area of 110 m$^2$.  The focal plane instrument is made up of 499 photo-multiplier tubes (PMTs) each with a 0.15$^\circ$ field of view (FoV), yielding a total camera FoV of 3.5$^\circ$.  The stereoscopic four-telescope array-system began operation in September of 2007.

VERITAS observed PG 1553+113  (VER J1555+111) from May 2010 to June 2012 for a total of 95 hours.  Observations were performed in wobble mode, with the source position offset from the telescope pointing direction by $0.5^\circ$, allowing for simultaneous background estimation.  The range of zenith angles for these observations was $20^\circ$ to $30^\circ$, with an average of $23^\circ$. The small zenith angles and event selection cuts optimized for a soft-spectrum source yield an analysis energy threshold  (energy of peak photon rate after cuts) of $180\,$GeV.  Events were reconstructed following the procedure outlined in \citet{Acciari:2008}.

The circular signal region used in the analysis was centered on the nominal source position and extended radially outward $0.14^\circ$.
  After applying quality selection criteria based on weather and instrument stability, and correcting for instrument read-out dead time, a total of 80 hours of live time were obtained. These   PG 1553+113 data yield an overall  detection significance of $53\sigma$   using Equation 17 of \citet{LiMa:1983}.     The excess is consistent with a point source.  The annual and cumulative analysis results are summarized in Table \ref{tab:ObsSummary}. 
\begin{table*}
	\caption{Summary of VERITAS observations from 2010, 2011, and 2012. The integral flux is denoted as $\Phi$ in the following.    i. The excess has been calculated using a normalization factor for the background of $\rm  \alpha = 0.143$.}
		\centering	
	\begin{tabular}{c c c c c c c c c}
		\hline\hline 
		& Live & Signif.  & On & OFF &  $\rm Excess^{i}$ &  $\Gamma$ & $\Phi(>200 \, \mathrm{GeV})$ & \% Crab \\ 
		& Time &           &      &       &    &  &              &  Nebula\\ 
		& $\left[ \mathrm{hours} \right]$ & $\left[ \sigma \right]$ &   &  & &  ($dN/dE \propto E^{-\Gamma}$) & $\left[ 10^{-11} \, \mathrm{cm}^{-2} \, \mathrm{s}^{-1} \right]$   &  Flux \\
		\hline
		2010 & 25 & 27 &   4,800  &   22,000 &  1,654 &  $4.37 \pm 0.16$ & $1.64 \pm 0.11$ & 6.8 \\ 
		2011 & 39 &    31&   6,490 &  30,400  &  2,143 & $4.35 \pm 0.15$ & $1.35 \pm 0.08$ & 5.5 \\ 
		2012 & 16 & 36 &   3,250 &  11,100  &  1,663 &   $4.28 \pm 0.14$ & $2.50 \pm 0.14$ & 10.3 \\ 
		\hline
		Total & 80 & 53 &   14,540 &   63,500  &  5,460&  $4.33 \pm 0.09$ & $1.69 \pm 0.06$ & 6.9 \\ 
		\hline\hline
	\end{tabular}
\label{tab:ObsSummary}
\end{table*}

\begin{figure}[t] 
        \centering
            \includegraphics[width=5in]{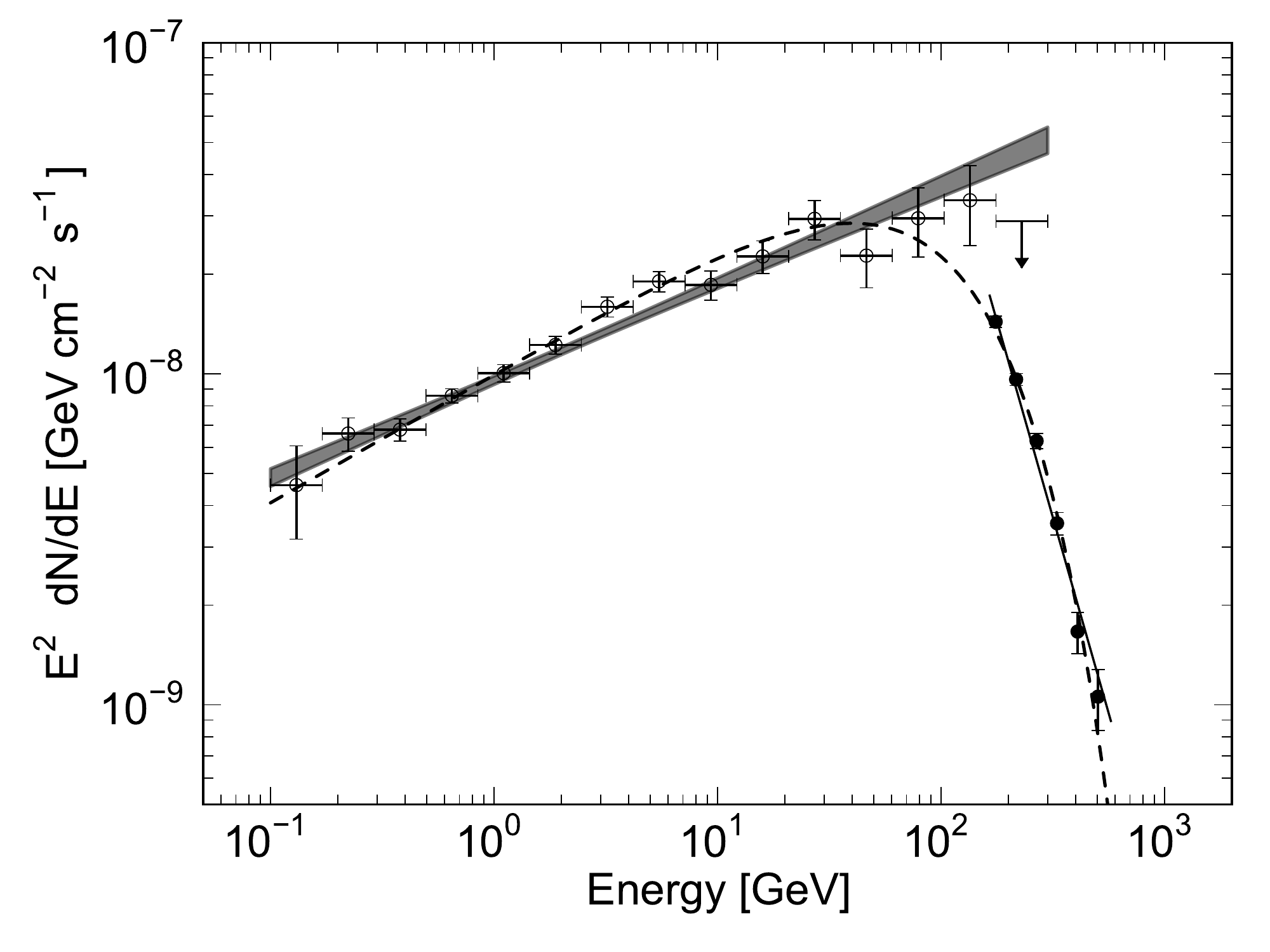}
	\caption{\textit{Fermi}-LAT spectrum of PG 1553+113 (grey shaded area and open data points) plotted along with the VERITAS spectrum (solid black data points and line).  The highest energy bin in the \textit{Fermi}-LAT spectrum represents the 95\% confidence level upper limit of the flux in this bin.  The dashed lines shows the best fit to the combined spectrum using a power law with an exponential cutoff.}
	\label{fig:SpectrumGeVTeV} 
\end{figure}

Figure \ref{fig:SpectrumGeVTeV} shows the time-averaged VERITAS spectrum for PG 1553+113 (black data points).  The spectrum is well fit by a power law of the form:
\begin{equation}
	\left( \frac{dN}{dE} \right)_{\it VERITAS} = (4.80 \pm 0.17) \times 10^{-11} \left( \frac{E}{0.3\,\mathrm{TeV}} \right)^{-4.33 \pm 0.09} \, \mathrm{ph} \, \mathrm{cm}^{-2} \, \mathrm{s}^{-1} \, \mathrm{TeV^{-1}}.
\end{equation}

\noindent yielding a fit probability of 3\% ($\chi^2/\nu = 10.8/4$).
The time-averaged integral flux above 200 GeV is $(1.69 \pm 0.06) \times 10^{-11} \, \mathrm{ph} \, \mathrm{cm}^{-2} \, \mathrm{s}^{-1}$, or 6.9\% of the Crab Nebula flux \citep{Mohanty:1998}.  Extensive studies of systematic uncertainties of the spectral index were performed for a range of sources with soft and hard spectral indices \citep{Madhavan:2013} by varying cut efficiencies, indicating that the systematic uncertainties of the  spectral index is less than 0.2.   Systematic uncertainties associated with the absolute energy calibration due to throughput uncertainties are estimated at the level of 20\%, thereby also causing systematic uncertainties of 55\% for the absolute fluxes for the very soft spectrum of PG 1553+13.   A secondary analysis of PG 1553+113 using an independent analysis package, yields an energy spectrum that is within these systematic uncertainties.

\section{\textit{Fermi}-LAT Observations}
\label{sec:FermiObs}
In addition to being a strongly-detected source at VHE, PG 1553+113 is also bright in the HE regime.  A total of 1227 days ($\sim \negthinspace 3.4\,$years) of LAT observations yield a detection significance of $81\sigma$.  The source is one of only 104 blazars (58 flat-spectrum radio quasars, 42 BL Lacs, and 4 with unknown classification) in the LAT Bright AGN Sample (LBAS) obtained from the first three months of \textit{Fermi}-LAT data \citep{Abdo:2009}.

The analysis of \textit{Fermi}-LAT data for PG 1553+113 was performed using the Pass 7 version of photon selection and the \texttt{ScienceTools-v9r23p1} suite of analysis tools.  Data were extracted within a $40^\circ \times 40^\circ$ square on the sky centered around the PG 1553+113 source position and binned into $0.2^\circ \times 0.2^\circ$ pixels using an Aitoff projection.   Events with energies between $100\,$MeV and $300\,$GeV were selected and binned in energy using 10 bins per decade.  A binned likelihood analysis was performed on the resulting photon counts cube.  

The best-fit spectrum obtained using the binned likelihood analysis is given by:
\begin{equation}
	\left( \frac{dN}{dE} \right)_{\it Fermi} = (2.42 \pm 0.06) \times 10^{-12} \left( \frac{E}{2239\,\mathrm{MeV}} \right)^{-1.71 \pm 0.02} \, \mathrm{ph} \, \mathrm{cm}^{-2} \, \mathrm{s}^{-1} \, \mathrm{MeV^{-1}},
\end{equation}
where $2239\,$MeV is the de-correlation energy quoted in the \textit{Fermi}-LAT Second Source Catalog \citep{Nolan:2012}.  Figure \ref{fig:SpectrumGeVTeV} shows the butterfly fit obtained from the \textit{Fermi}-LAT data (grey shaded area) along with individual flux points (open circles).  The flux points were calculated using the best-fit spectral index and fitting the flux normalization independently within each energy bin.  The spectral index is in good agreement with that from the  \textit{Fermi}-LAT 2 Year Catalog ($1.67 \pm 0.02$) \citep{Nolan:2012}.
As can be seen in Figure \ref{fig:SpectrumGeVTeV}, the transition of the spectrum from the HE to VHE regime is very sharp.  This sharp change in spectral slope could be predominantly a result of extragalactic background light (EBL) absorption given that PG 1553+113 is known to have a redshift of at least 0.395 \citep{Danforth:2010}.

The combined \textit{Fermi}-LAT and VERITAS spectra show clear evidence of a cutoff at $\sim \negthinspace 100\,$GeV. The best-fit parameters resulting from a fit to a power law with an exponential cutoff are:
\begin{equation}
	\left( \frac{dN}{dE} \right)_{\it combined} = (2.46 \pm 0.08) \times 10^{-10} \left( \frac{E}{10\,\mathrm{GeV}} \right)^{-1.61 \pm 0.02} \exp \left( \frac{-E}{(101.9 \pm 3.2)\,\mathrm{GeV}} \right) \, \mathrm{ph} \, \mathrm{cm}^{-2} \,\mathrm{s}^{-1} \, \mathrm{GeV^{-1}},
\end{equation}
yielding a fit probability of 39\% ($\chi^2/\nu = 18/17$).\footnote{The fit uses the \textit{Fermi}-LAT flux points but does not include the upper limit between 176 and 223\,GeV.}  Attempting to fit the spectrum with a log-parabolic function does not give an acceptable fit ($\chi^2/\nu = 456/17$).

\section{Multiwavelength Variability}
\label{sec:Variability}
\begin{figure}[t]
	\centering
	\includegraphics[width=4.95in, angle=-90.0]{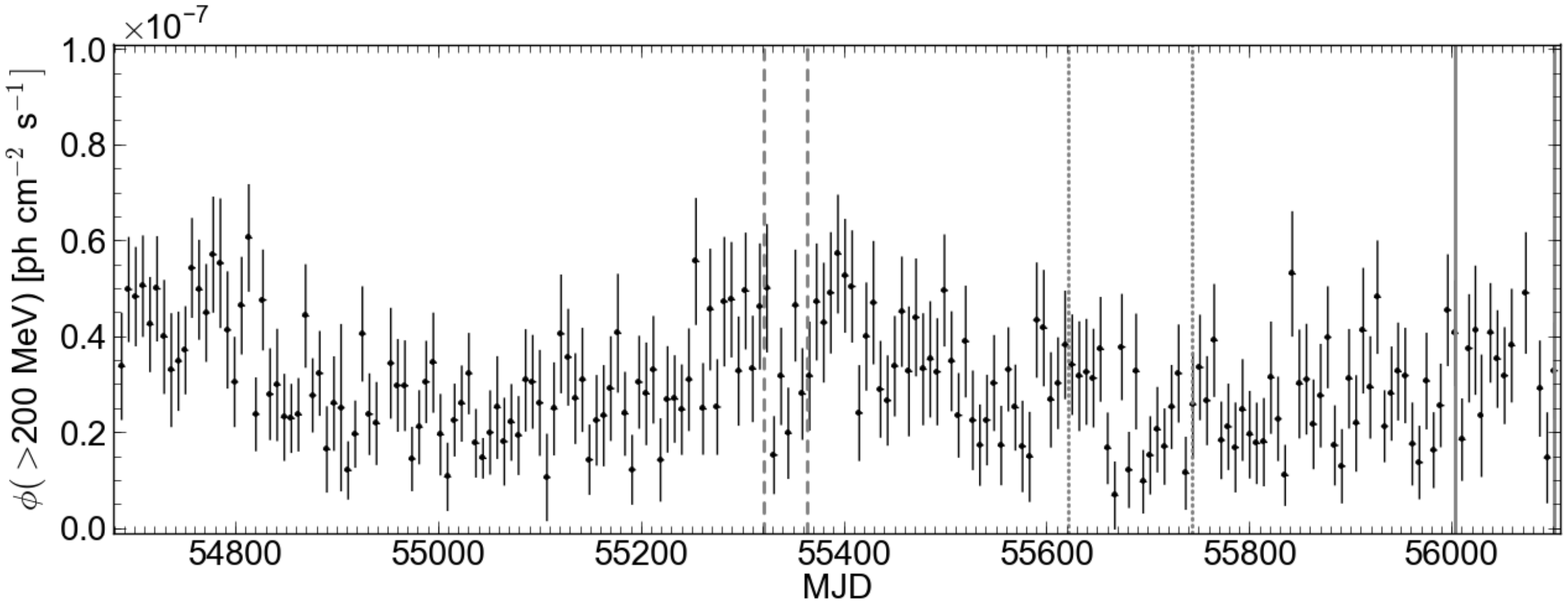}
	\caption{\textit{Fermi}-LAT PG 1553+113 weekly integral flux lightcuve above 200 MeV.  The grey dashed, dotted, and solid lines indicate the time periods of VERITAS observations during 2010 (May 4 - June 17), 2011 (February 28 - July 4), and 2012 (March 15 - June 24), respectively. The detailed VERITAS light curves and the corresponding observing dates can be seen from Figures \ref{fig:FermiVERITASLightCurves2010}, \ref{fig:FermiVERITASLightCurves2011} and \ref{fig:SwiftFermiVERITASLightCurves2012}. }
	\label{fig:FermiLightCurve}
\end{figure}

MAGIC observations of PG 1553+113, spread over five years (2005-2009), indicate flux variability above 150\,GeV \citep{Aleksic:2012}.  The lowest and highest flux states measured by MAGIC during this time period differ by a factor of $\sim \negthinspace 3$.  Observations made by \textit{Fermi}-LAT exhibit clear signs of long-term variability above 200\,MeV, with a constant-flux probability of $\sim \negthinspace  3 \times 10^{-6}$. This is illustrated in Figure \ref{fig:FermiLightCurve} where the integral flux above $200\,$MeV, binned weekly, is shown.  This variability occurs on approximate timescales of hundreds of days.  In the analysis of the first $\sim \negthinspace 200$ days of \textit{Fermi}-LAT data, \citet{Abdo:2010:PG1553} report a constant flux ($E \negthinspace > \negthinspace 200\,$MeV) fit probability of 54\% using a two-day binning timescale.

\begin{figure}[t]
	\centering
	\includegraphics[width=5in]{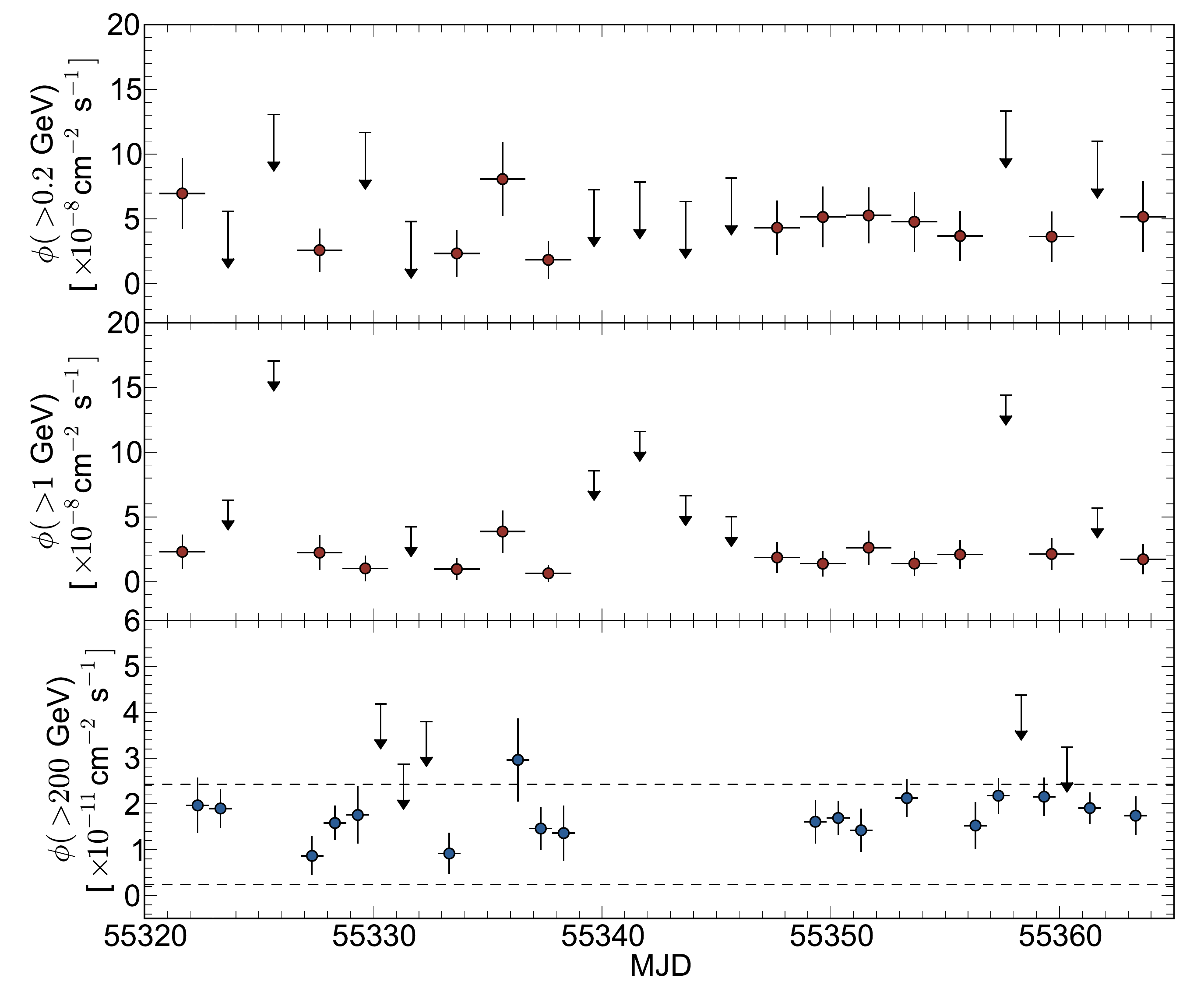}
	\caption{\textit{Fermi}-LAT PG 1553+113 bi-daily flux light curve (2010) above 200\,MeV (upper plot) and above 1\,GeV (middle plot), in units of $\mathrm{cm}^{-2} \, \mathrm{s}^{-1}$, for the time periods contemporaneous with VERITAS observations and VERITAS daily integral flux light curve above 200 GeV (lower plot), in units of $\mathrm{cm}^{-2} \, \mathrm{s}^{-1}$ (note, for space reasons we do not use notation of ph~$\mathrm{cm}^{-2} \, \mathrm{s}^{-1}$).  The upper and lower dashed lines indicate the integral fluxes above 200\,GeV corresponding to 10\% and 1\% of the Crab Nebula flux, respectively.  The black arrows in all light curves represent $2\sigma$ upper limits.}
	\label{fig:FermiVERITASLightCurves2010}
\end{figure}

Figures \ref{fig:FermiVERITASLightCurves2010}  and \ref{fig:FermiVERITASLightCurves2011} show the contemporaneous \textit{Fermi}-LAT and VERITAS light curves for 2010 and 2011, respectively.  The \textit{Fermi}-LAT data are binned into two-day-wide bins and the VERITAS data into one-day-wide bins.  The LAT-measured integral fluxes above $200\,$MeV and above $1\,$GeV are shown along with the VERITAS-measured integral flux above $200\,$GeV.  The average integral fluxes above $200\,$GeV during the 2010 and 2011 VERITAS observing seasons were 6.8\% and 5.5\% of the Crab Nebula flux, respectively.  Both the \textit{Fermi}-LAT and VERITAS data are consistent with resulting from a steady flux over these time periods. 

\begin{figure}[t]
	\centering
	\includegraphics[width=5in]{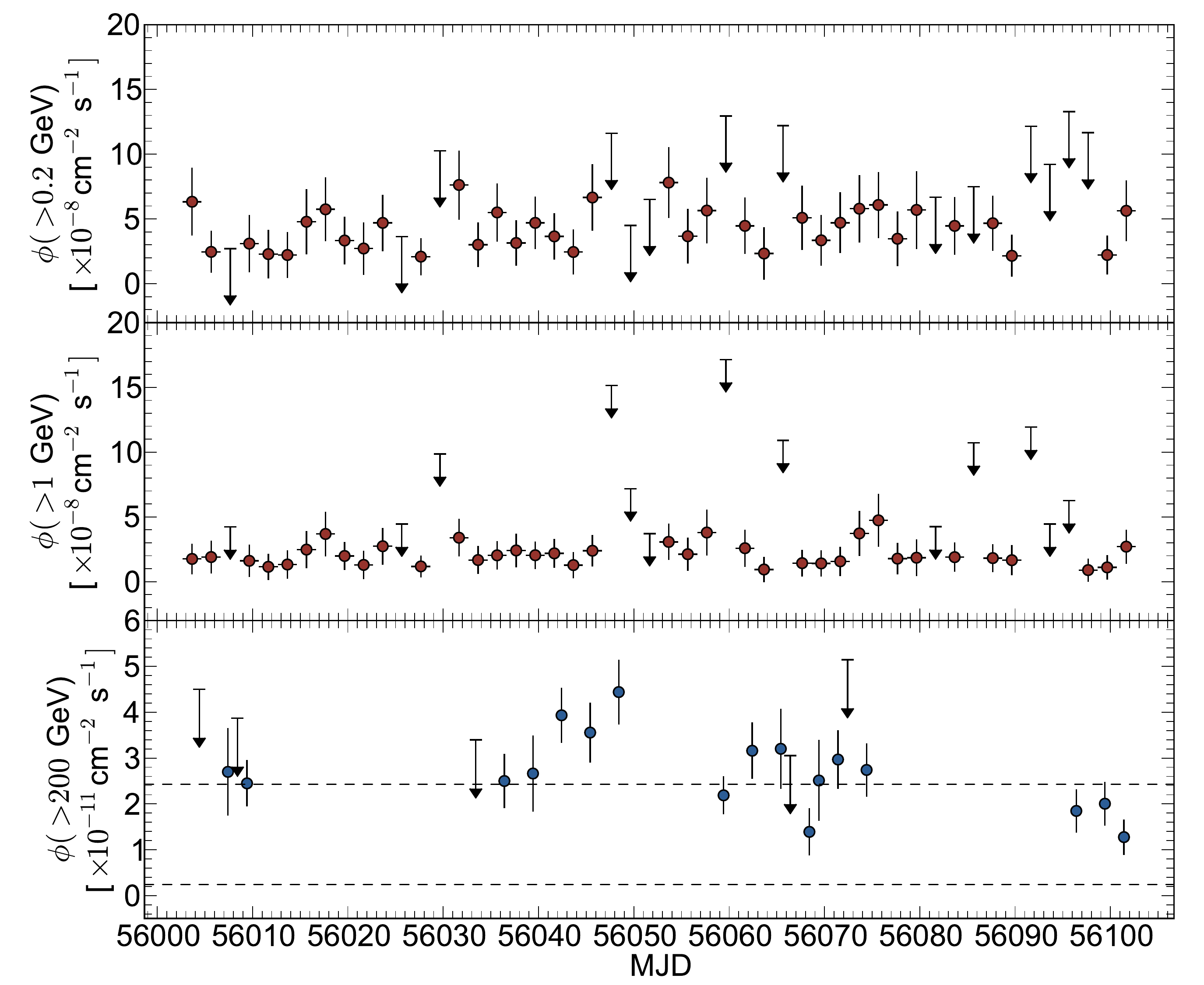}
	\caption{\textit{Fermi}-LAT PG 1553+113 bi-daily flux light curve (2012) above 200\,MeV (upper plot) and 1\,GeV (middle plot), in units of $\mathrm{cm}^{-2} \, \mathrm{s}^{-1}$, for the time periods contemporaneous with VERITAS observations, and VERITAS daily integral flux light curve above 200 GeV (lower plot), in units of $\mathrm{cm}^{-2} \, \mathrm{s}^{-1}$.  The upper and lower dashed lines indicate the integral fluxes above 200\,GeV corresponding to 10\% and 1\% of the Crab Nebula flux, respectively.  The black arrows represent $2\sigma$ upper limits.}
	\label{fig:SwiftFermiVERITASLightCurves2012}
\end{figure}

Figure \ref{fig:SwiftFermiVERITASLightCurves2012} shows the LAT and VERITAS light curves for 2012.  The average integral flux above $200\,$GeV in 2012 was 10.2\% of the Crab Nebula flux.  Marginal flux variability in 2012 is suggested, since the  probability that the VERITAS data result from a steady source flux is just 0.03\%. On  the other hand, \textit{Fermi}-LAT data during the time periods of  VERITAS observations show no evidence for variability, as is indicated by large probabilities for a steady flux measured by applying a $\rm \chi^{2}$ test.   The variability results for \textit{Fermi}-LAT and VERITAS from 2010, 2011, and 2012 are summarized in Table \ref{tab:MultiWavelengthSummary}.

The average flux measured by VERITAS during the 2012 observing season is clearly elevated with respect to the fluxes from the 2010 and 2011 observations.  During 2012, the flux of PG~1553+113 reached 18\% of the Crab Nebula flux ($> \negthinspace 200\,$GeV) -- approximately a factor of 3 above the average flux between 2010 and 2011.  Taking the full data set into consideration, the fit probability for a steady flux over the full three years of observations is $\sim 10^{-6}$.    This provides strong evidence that PG 1553+113 is variable over timescales on the order of years.  Similarly, observations by the  MAGIC collaboration \citep{Aleksic:2012} show significant flux variations during the time period 2007, 2008 and 2009, with a high state during 2008.   However, results reported from VERITAS observations also include contemporaneous \textit{Fermi}-LAT  fluxes, which do not exhibit flux variations at a similar level.  It should be noted that despite the fact that the \textit{Fermi}-LAT fluxes have only slightly larger statistical uncertainties (\textit{Fermi}-LAT: 10 - 18\% vs. VERITAS: 5 - 6\%), contemporaneous \textit{Fermi}-LAT flux variations are not suggested by the data.

\begin{figure}[t]
	\centering
	\includegraphics[width=5in]{Fig4.pdf}
	\caption{\textit{Fermi}-LAT PG 1553+113 bi-daily flux light curve (2011) above 200\,MeV (upper plot) and 1\,GeV (middle plot), in units of $\mathrm{cm}^{-2} \, \mathrm{s}^{-1}$, for the time periods contemporaneous with VERITAS observations, and VERITAS daily integral flux light curve above 200 GeV (lower plot), in units of $\mathrm{cm}^{-2} \, \mathrm{s}^{-1}$.  The upper and lower dashed lines indicate the integral fluxes above 200\,GeV corresponding to 10\% and 1\% of the Crab Nebula flux, respectively.  The black arrows represent $2\sigma$ upper limits.}
	\label{fig:FermiVERITASLightCurves2011}
\end{figure}

\begin{table*}
	\caption{Summary of average fluxes from contemporaneous time periods with  \textit{Fermi}-LAT and VERITAS data in 2010, 2011 and 2012.}
	\centering	
	\begin{tabular}{c | c | c | c | c | c | c}
		\hline\hline 
		& \multicolumn{4}{c|}{\textit{Fermi}-LAT} & \multicolumn{2}{c}{VERITAS} \\
		\cline{2-7}
		& $ \Phi (\geq 0.2 \, \mathrm{GeV}) $ & Steady Flux   & $     \Phi (\geq 1 \, \mathrm{GeV}) $ & Steady Flux   & $   \Phi (\geq 200 \, \mathrm{GeV}) $ & Steady Flux  \\
		Year & $\times 10^{-8} \,   \mathrm{cm}^{-2} \, \mathrm{s}^{-1}$ &  Prob. [\%]   &    $\times 10^{-8} \,  \mathrm{cm}^{-2} \, \mathrm{s}^{-1}$ &   Prob. [\%] & $\times 10^{-11} \, \mathrm{cm}^{-2} \, \mathrm{s}^{-1}$  & Prob. [\%] \\
		\hline
		2010 & $3.84 \pm 0.59$ &  62.5  &   $1.39 \pm 0.26$ & 77.9 & $1.70 \pm 0.11$   &  37.1 \\
		2011 & $2.92 \pm 0.32$ &   77.7 &    $1.42 \pm 0.16$ &99.7  &   $1.54 \pm 0.08$  & 6.4 \\
		2012 & $3.77 \pm 0.34$ & 96.2   &  $1.83 \pm 0.19$ & 99.8   &  $2.42 \pm 0.14$   &  0.03 \\
		\hline\hline
	\end{tabular}
\label{tab:MultiWavelengthSummary}
\end{table*}

\section{Constraining the Redshift of PG 1553+113}
\label{sec:Redshift}
To date, no attempt at measuring the redshift of PG 1553+113 has been successful due to its featureless optical spectrum.   The first measurement was attempted by \citet{Miller:1983}   using the \textit{International Ultraviolet Explorer},  who reported the redshift to be $z=0.36$.  This measurement was later disputed as a spurious emission line from instrumental effects, misidentified as a Lyman-$\alpha$ (Ly$\alpha$) line at $z=0.36$ \citep{Falomo:1990}.  However, the use of this incorrect value can still be found in current literature.   A variety of limits have been obtained using both optical and VHE observations (Table \ref{tab:RedshiftSummary}).  The two approaches are complementary to one another as the techniques utilizing optical spectra provide (in general) lower limits on the source redshift whereas VHE observations can be used to obtain upper limits.  The following two subsections summarize the methods used to constrain the redshift of PG 1553+113.

\subsection{Optically-Based Redshift Limits}
The Wide Field Planetary Camera 2 (WFPC2), on board the \textit{Hubble Space Telescope} (HST), was used to survey 132 BL Lac Objects \citep{Urry:2000}.  This survey demonstrated that, apart from having highly-active nuclei, these objects appear to be typically hosted by elliptical galaxies.  The distribution of R-band absolute magnitudes ($M_R$) for the host galaxies is well fit by a Gaussian distribution with a mean $M_R = -22.8$ and standard deviation $\sigma = 0.5$ \citep{Sbarufatti:2005:ImagingRedshifts}.  Given their intrinsic similarities, the apparent magnitudes ($m_R$) of BL Lac host galaxies can be used to obtain their redshifts.  The average difference between these photometric redshift measurements and those obtained spectroscopically is $\Delta z = 0.01 \pm 0.05 \, \mathrm{(rms)}$ \citep{Sbarufatti:2005:ImagingRedshifts}.  

If the host galaxy is not detected, then a lower limit on $m_R$ (upper limit on the luminosity) can be used to obtain a lower limit on the source redshift.  This technique was used by \citet{Sbarufatti:2005:ImagingRedshifts} who calculated a lower limit on the host galaxy apparent magnitude of PG 1553+113 of $m_R \negthinspace > \negthinspace 21.6$, corresponding to a redshift lower limit of $z \negthinspace > \negthinspace 0.78$.  In deriving their limit on $m_R$, \citet{Sbarufatti:2005:ImagingRedshifts} assumed the dominant source of error was statistical.  This was later revisited by \citet{Treves:2007} who showed that systematic effects cannot be ignored.  Taking this into consideration, \citet{Treves:2007} calculated a lower limit on the apparent magnitude of $m_R \negthinspace > \negthinspace 18.07$ and a redshift lower limit of $z \negthinspace > \negthinspace 0.25$.

Another approach to constraining BL Lac redshifts, which also utilizes the standard-candle nature of the host galaxies, assumes that the optical spectrum of the BL Lac object consists of two components -- nonthermal emission from the nucleus distributed as a power law and a thermal component from the host galaxy \citep{Sbarufatti:2006,Finke:2008}.  Depending on the nucleus-to-host flux ratio, the optical spectrum may be dominated by the featureless nonthermal emission from the nucleus or by the spectral signature of the host galaxy.  This ratio is therefore related to the equivalent width of a given spectral absorption line.  The nucleus-to-host flux ratio is also related to the apparent magnitude of the nucleus through the assumption that the absolute magnitude of the host galaxy can be approximated using a mean value as discussed above (i.e., $M_R \approx -22.8$).

In the absence of detectable absorption or emission lines from optical observations of the host galaxy, the minimum detectable equivalent width ($\mathrm{EW}_\mathrm{min}$) is used to constrain the source redshift.  The expected redshift dependence of the nucleus-to-host flux ratio at a fixed wavelength differs when using $\mathrm{EW}_\mathrm{min}$ or the apparent magnitude of the nucleus.  The nucleus-to-host flux ratio calculated at a particular redshift using the apparent magnitude must be greater than or equal to the flux ratio calculated at the same redshift using $\mathrm{EW}_\mathrm{min}$.  If this were not the case, the equivalent width of a given spectral feature in the host galaxy would exceed $\mathrm{EW}_\mathrm{min}$, thereby making it detectable.  Using this technique, \citet{Sbarufatti:2006} placed a lower limit on the redshift of PG 1553+113 of $z \negthinspace > \negthinspace 0.09$.

Interstellar (in the host galaxy) and intergalactic absorption features present in spectra can also also be used to place lower limits on source redshifts.  This technique was recently performed on PG 1553+113 observations using the \textit{Hubble Space Telescope}  COS~\citep{Danforth:2010},  utilizing spectral absorption features over the wavelength range $1135 \, \mathrm{\AA} \negthinspace < \negthinspace \lambda \negthinspace < \negthinspace 1795 \, \mathrm{\AA}$.  Based on a Ly$\alpha$+O VI absorber, \citet{Danforth:2010} find a lower limit of $z \negthinspace > \negthinspace 0.395$, with a somewhat larger lower limit  of $z \negthinspace > \negthinspace 0.433$ being found from a single weak Ly$\alpha$ line detection.  The existing COS data are sensitive to Ly$\alpha$ absorbers with redshifts $z \negthinspace < \negthinspace 0.47$.  However, \citet{Danforth:2010} present statistical arguments for a $1\sigma$ upper limit on the redshift of $z \negthinspace \leq \negthinspace 0.58$ based on the lack of detection of Ly$\beta$ lines at redshifts $z \negthinspace > \negthinspace 0.4$.  Assuming the validity of these arguments, the most current constraints on PG 1553+113 place its redshift in the range of  $0.395 \negthinspace < \negthinspace  z \negthinspace \lesssim \negthinspace 0.58$.

\subsection{VHE-Based Redshift Limits}
The use of VHE gamma-ray spectra to constrain the redshifts of blazars exploits the fact that VHE gamma rays, as they traverse cosmological distances, may produce $e^+ e^-$ pairs through their interaction with the diffuse infrared to ultraviolet wavelength photons of the EBL.  The amount of VHE gamma-ray absorption depends on the redshift of the source in question and the spectral energy distribution (SED) of the EBL.  For blazars with well-measured redshifts, the intrinsic VHE spectrum can be calculated by assuming an EBL scenario, calculating the $\gamma_{VHE} \, \gamma_{EBL}$ optical depth, $\tau(E_\gamma)$, as a function of gamma-ray energy, $E_\gamma$, and then applying a correction factor of $e^{\tau(E_\gamma)}$ to the observed flux in each energy bin.  

For blazars with unknown redshifts, the optical depth over a range of redshifts can be calculated using a background-light SED constituting a lower limit on the EBL.  This in turn provides a lower limit on the gamma-ray absorption at each redshift.  The absorption increases  with both redshift and gamma-ray energy.  For sufficiently high redshifts, the calculated intrinsic VHE spectrum may take on an unphysical shape.  Namely, it may exhibit a statistically significant exponential rise in flux with energy  \citep{Dwek:2005},  as determined by using the F-test to calculate the probability that a reduction in chi-square of the fit due to the inclusion of an exponential rise with energy exceeds the value which can be attributed to random fluctuations in the data (denoted as Exponential Rise in Table 3).  Such spectral features are inconsistent with standard synchrotron self-Compton \citep{Maraschi:1992,Bloom:1996} and external inverse-Compton \citep{Dermer:1993,Sikora:1994} models of blazars.  It can therefore be concluded that the calculated gamma-ray absorption in these cases is too large and, consequently, the assumed redshift must be too large.  In this way, an upper limit on the redshift of a particular VHE blazar can be obtained.

Another simpler approach is to calculate the de-absorbed spectrum and place a cut on the hardest (minimum) ``acceptable" value for the intrinsic VHE spectral index, thereby constraining the maximum redshift.  This value can be motivated by theoretical arguments \citep{Aharonian:2006:Nature} (e.g., $\Gamma_\mathrm{int} \negthinspace < \negthinspace 1.5$) or from observations of the source at lower energies where EBL attenuation is negligible \citep{Georganopoulos:2010} (e.g., $\Gamma_\mathrm{int} \negthinspace < \negthinspace \Gamma_\mathrm{LAT}$).

\begin{table}[t]
	\centering
	\caption{Summary of redshift constraints for PG 1553+113.  \textit{Column 1:} Observation waveband used for redshift constraint.  \textit{Column 2:} Technique used for redshift constraint. \textit{Column 3:} Calculated redshift limit.  \textit{Column 4:} Journal reference.}
	\begin{tabular}{c c c c}
		\hline\hline
              Waveband & Technique & Redshift Limit & Reference \\
		\hline\
		Optical & $m_R > 21.60$ & $0.78 < z \,\quad\quad\quad$ & \citet{Sbarufatti:2005:ImagingRedshifts} \\
		VHE & $\Gamma_\mathrm{int} > 1.5$  & $\quad\quad\quad z < 0.74$ & \citet{Aharonian:2006} \\
		Optical & EW$_\mathrm{min}$ & $0.09 < z \,\quad\quad\quad$ & \citet{Sbarufatti:2006} \\
		VHE & $\Gamma_\mathrm{int} > 1.5$  & $\quad\quad\quad z < 0.74$ & \citet{Albert:2007} \\
		VHE & $\Gamma_\mathrm{int} > 1.5$ & $\quad\quad\quad z < 0.69$ & \citet{Mazin:2007} \\
              VHE & VHE Pileup & $\quad\quad\quad z < 0.42$ & \citet{Mazin:2007} \\
              Optical & $m_R > 18.07$ & $0.25 \leq z \,\quad\quad\quad$ & \citet{Treves:2007} \\
              VHE & $\Gamma_\mathrm{int} > 1.5$  & $\quad\quad\quad z < 0.69$ & \citet{Aharonian:2008} \\
		VHE & $\Gamma_\mathrm{int} > 1.5$ & $\quad\quad\quad z < 0.67$ & \citet{Prandini:2009} \\
              VHE & VHE Pileup & $\quad\quad\quad z < 0.58$ & \citet{Prandini:2009} \\
              UV & IGM Absorp. & $0.395 < z  < 0.58$ & \citet{Danforth:2010} \\
              VHE & Exponential Rise & $\quad\quad\quad z < 0.62$ & This work (EBL evolution) \\
               VHE & Exponential Rise & $\quad\quad\quad z < 0.53$ & This work (no EBL evolution) \\
		\hline
	\end{tabular}
	\label{tab:RedshiftSummary}
\end{table}

The VERITAS spectrum presented in Figure \ref{fig:SpectrumGeVTeV} was used to place an upper limit on the redshift of PG 1553+113.  The technique used follows that described above in which the redshift is increased until a statistically significant exponential rise in the intrinsic spectrum is present.  The EBL SED  of \citet{Kneiske:2010} was used,  which reproduces the EBL flux lower limits obtained from galaxy counts.  As such, it represents a lower limit on the EBL and the minimum amount of EBL absorption for gamma rays.  The resulting upper limit  on the redshift is therefore conservative given that stronger EBL absorption will introduce an exponential rise in the intrinsic VHE spectrum at lower redshifts.  

\begin{figure}[t]
	\centering
	\includegraphics[width=5in]{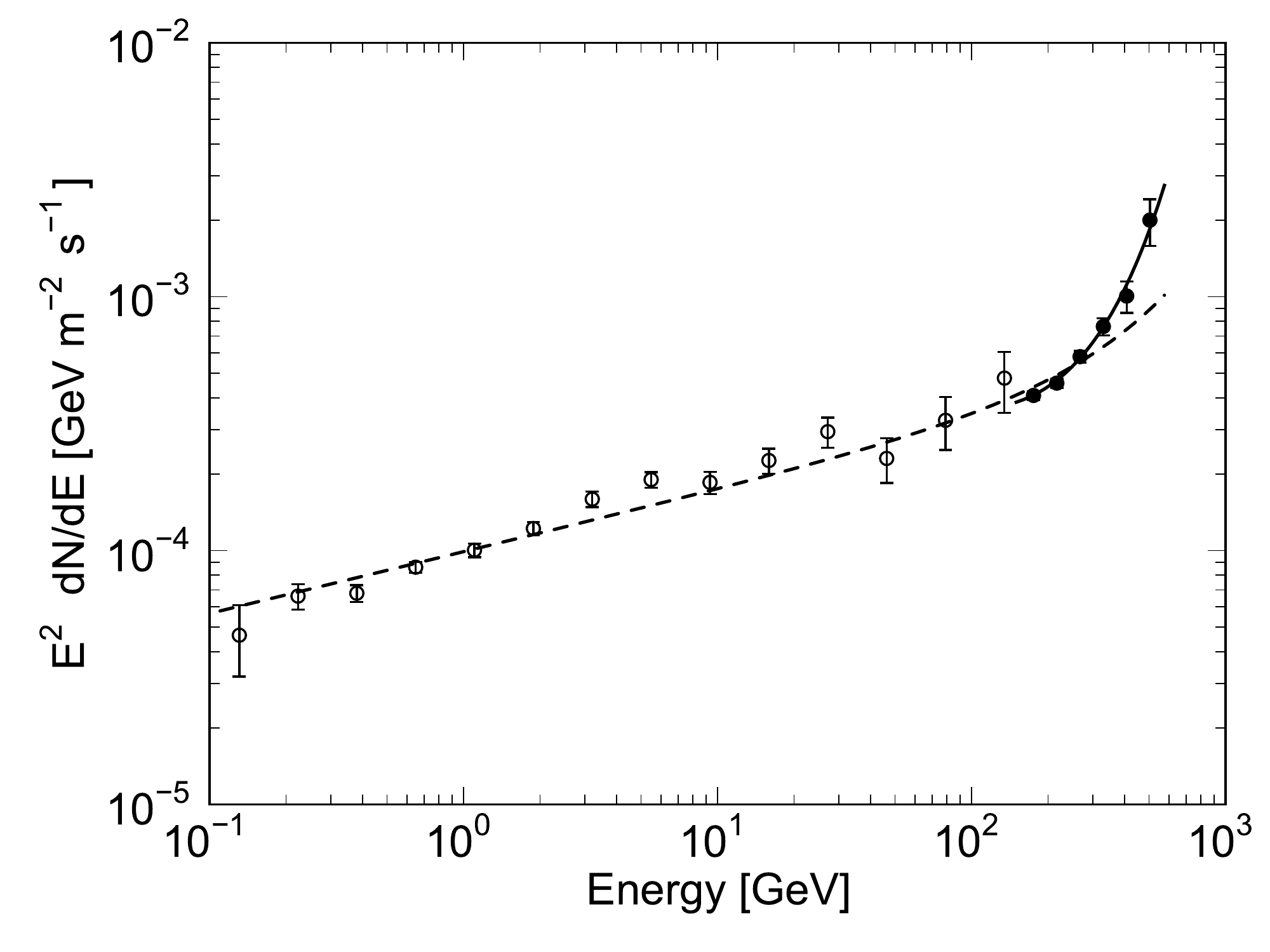}
	\caption{High-energy (open circle data points) and VHE (filled circle data points) intrinsic spectrum for PG 1553+113 assuming the EBL model of \citet{Kneiske:2010} and a redshift of $z=0.53$.  The solid curve represents the best fit to the intrinsic VHE spectrum using a power law with an exponential rise and was the fit used to set the upper limit on the source redshift.  The dashed curve shows the best fit to the intrinsic spectrum covering the full energy range of \textit{Fermi}-LAT and VERITAS.}
	\label{fig:SpectrumGeVTeVIntr}
\end{figure}

The statistical significance of the exponential rise in the intrinsic spectrum was calculated using the F-test, following the prescription described in \citet{Dwek:2005}.  The 95\% confidence level redshift upper limit obtained from the VERITAS spectrum is $z \negthinspace \leq \negthinspace 0.53$.  The resulting intrinsic spectrum for this maximum redshift is shown in Figure \ref{fig:SpectrumGeVTeVIntr}, along with the de-absorbed \textit{Fermi}-LAT spectrum.  Note that, while the \textit{Fermi}-LAT spectrum has been de-absorbed, it was not used to constrain the source redshift.  As can be seen from the fit to the full spectrum in Figure \ref{fig:SpectrumGeVTeVIntr} (dashed curve), a power law with an exponential rise is not sufficient to describe the quickly rising VHE flux over such a broad energy range.    Due to possible systematic uncertainties in the relative energy and flux calibration between the Fermi-LAT and the VERITAS data and the fact that the Fermi-LAT data are not strictly contemporaneous, we use only the VERITAS data for constraining the redshift.

There is one correction that must be applied to the redshift upper limit calculated above.  The EBL consists of the collective emission from galaxies and AGN over the history of the Universe.  As a result, the intensity and (to a lesser degree) the shape of the EBL evolve with redshift, reflecting the evolution of the source populations producing this cosmic radiation (e.g., \citet{Franceschini:2008}).  It is important to take into account the evolution of the EBL when calculating the absorption of gamma rays produced by distant sources.  Failure to account for EBL evolution will result in an overestimation of the gamma-ray attenuation -- reaching the $\sim  10\%$ level at a redshift of $z = 0.2$ \citep{Raue:2008}.  

In the approach of \citet{Kneiske:2010}, the evolution of the EBL is described using a simple stellar population model, dependent on different stellar masses, whose evolution is governed by the input comoving star formation rate density (with units of $\mathrm{M}_\odot \, \mathrm{Mpc}^{-3} \, \mathrm{yr}^{-1}$).  Reconstructed EBL SEDs at redshifts of $z=0.0, 0.1, 0.2, 0.3, 0.8, \mathrm{and} ~ 2.0$ are presented, but it is the EBL SED at $z=0$ that is most often used when calculating the gamma-ray optical depth.  Incorporating the evolutionary model of the EBL into the gamma-ray optical depth calculation requires a detailed knowledge of the developed EBL model as a function of redshift.

A simple approach incorporating the evolution of the EBL into the calculation of the gamma-ray optical depth is to introduce a correction factor into the EBL photon number density scaling with redshift.  If the sources contributing to the EBL were to completely shut off (i.e., no new photons were created) at a redshift, $z=z'$, the EBL photon number density would scale as $n(\epsilon,z) \propto (1+z)^3$, where $\epsilon$ is the EBL photon energy, between the redshifts of $z=0$ and $z=z'$.  When one calculates the gamma-ray opacity for a redshift of, e.g, $z = 0.1$, and ignores the evolution of the EBL, the inherent assumption is that there were no additional contributions to the EBL between $z = 0.1$ and $z = 0$, and the EBL photon number density at $z = 0.1$ is therefore $n(\epsilon,0.1) = n(\epsilon,0)(1+0.1)^3$.

If, however, one assumes that galaxies between redshifts of $z = 0.1$ and $z = 0$ are contributing to the EBL, the increase in the photon number density with redshift will scale more slowly than $(1+z)^3$.  To account for this, an evolutionary factor, $f_\mathrm{evo}$, can be introduced such that the EBL photon number density scales as $n(\epsilon,z) \propto (1+z)^{3-f_\mathrm{evo}}$.  The appropriate value for $f_\mathrm{evo}$ can be determined by comparing the predicted EBL evolution with more detailed evolutionary models (e.g., \citet{Kneiske:2002} and \citet{Primack:2005}).  \citet{Raue:2008} have shown that a number density scaling correction factor of $f_\mathrm{evo} = 1.2$ yields a redshift evolution in good agreement with the more detailed models of \citet{Kneiske:2002} and \citet{Primack:2005}, out to a redshift of $z \sim 0.7$.

When accounting for EBL evolution using the approach described above, the 95\% confidence level upper limit on the redshift of PG 1553+113 is $z \negthinspace \leq \negthinspace 0.62$. 

 In summary the redshift upper limit calculation given here is very conservative as it uses a much improved VHE spectrum of PG1553+113 and a method with minimal assumptions about the source spectrum.  Furthermore, we use the  minimally required EBL from \citet{Kneiske:2010}, consistent with lower limits from galaxy counts, whereas a previously published redshift 
upper limit from VHE observations used a higher EBL (see  \citet{Mazin:2007}).  And finally,  we show the effects of potential EBL evolution on the redshift upper limit.

\section{Discussion \& Conclusions}
\label{sec:Discussion}
PG 1553+113 was observed by VERITAS between May 2010 and June 2012 for a total of 95 hours resulting in a lifetime of 80 hours.  The time-averaged flux measured in 2012 was elevated with respect to the fluxes from 2010 and 2011 by a factor of 1.5 and 1.9, respectively.  There is evidence for VHE variability within the 2012 observing season, with the integral flux above 200\,GeV reaching as high as 18\% of the Crab Nebula flux.  The fluxes measured by \textit{Fermi}-LAT above 200\,MeV and 1\,GeV show no evidence of variability over the time periods contemporaneous with VERITAS observations.

The reconstructed VERITAS spectrum is soft -- with a spectral index of $\Gamma =  4.33 \pm 0.09$ -- while the spectrum measured by the \textit{Fermi}-LAT is hard -- having a spectral index of $\Gamma = 1.71 \pm 0.02$.  The combined spectrum is well fit by a power law with an exponential cutoff whose spectral index and cutoff energy are $\Gamma = 1.61 \pm 0.02$ and $E_\mathrm{cut} = 101.9 \pm 3.2$~GeV, respectively.  

The allowable redshift range of PG 1553+113 has narrowed considerably over the last several years, and with these new results presented here using the VERITAS energy spectrum combined with most conservative 
assumptions about the EBL and the intrinsic source spectrum, robust upper limits are now available.  Neglecting EBL evolution, the limits obtained from this work place PG 1553+113 at a redshift of $z \negthinspace \leq \negthinspace 0.53$.  When EBL evolution is included, the redshift upper limits could be as high as $z \negthinspace \leq \negthinspace 0.62$.  Including the lower limit from \citet{Danforth:2010} yields an allowable redshift range of $0.4 \negthinspace \lesssim \negthinspace z \negthinspace \lesssim \negthinspace 0.6$.  If a definitive measurement for the redshift of PG 1553+113 is obtained, and the value indeed turns out to be as large as expected, the source will prove to be an important probe for studying the EBL and its evolution.

\acknowledgements
The VERITAS Collaboration is grateful to Trevor Weekes for his seminal contributions and leadership in the field of VHE gamma-ray astrophysics, which 
made this study possible.  This research is supported by grants from the U.S. Department of Energy Office
of Science, the U.S. National Science Foundation and the Smithsonian Institution, by NSERC in Canada, by Science Foundation Ireland (SFI 10/RFP/AST2748) and by STFC in the U.K.   We acknowledge the excellent work of the technical support staff at the Fred Lawrence Whipple Observatory and at the collaborating institutions in the construction and operation of the instrument.

\bibliographystyle{apj}
\bibliography{apj-jour,bibliography}

\end{document}